\documentclass[twocolumn]{revtex4}
\usepackage[dvips]{graphicx}
\usepackage{amsfonts,amsmath}
\usepackage{float,color,array,multirow}

\begin{document}

\title{Magnetic field effects on spin-split band and magnon transport in altermagnets and emergent compensated ferrimagnets}

\author{Kazushi Aoyama$^1$ and Hikaru Kawamura$^2$}

\date{\today}

\affiliation{ $^1$Graduate School of Advanced Science and Engineering, Hiroshima University, Higashihiroshima 739-8521, Japan \\
$^2$Center for Life Photonic Innovation, Kobe University, Kobe 657-8501, Japan}

\begin{abstract}
In altermagnets and fully compensated ferrimagnets, not only the electron band but also the magnon band exhibits spin splitting without net magnetization, which enables thermal activation of the magnon spin current. Here, we theoretically investigate magnetic field effects on the magnon properties of these antiferromagnets in the presence of a weak easy-axis anisotropy which makes the collinear states robust against the magnetic field. 
For the altermagnet and compensated ferrimagnet, we analyze a 2 sublattice order in the $J_1$-$J_2$-$J_2^\prime$ model on the square lattice and a triple-${\bf Q}$ 12-sublattice order in the $J_1$-$J_3$ model on the kagome lattice, respectively, each accompanied by $d$-wave and $s$-wave spin splitting at zero field. 
It is shown that for positive (negative) magnetic field $H$ whose energy scale is smaller than the anisotropy gap, the up- and down-spin magnon bands are shifted to lower (higher) and higher (lower) energies, respectively, similarly to the Zeeman coupling in electron systems. In the altermagnet, with increasing field, the $d$-wave splitting tends to be deformed into the $s$-wave one, which is reflected as the change in the direction of the spin current generated by thermal gradient. In the compensated ferrimagnet, the $s$-wave nature, i.e., the population imbalance between the up- and down-spin magnons at $H=0$, can induce an asymmetric field dependence of the longitudinal spin and thermal conductivities in a field-sweep process. 
\end{abstract}

\maketitle
\section{Introduction}
An altermagnet, a third type of collinear magnetic state distinct from conventional ferromagnets and antiferromagnets, has attracted growing attention for its potential applications to spintronics devices \cite{alter_Smejkal_prx_22_overview, alter_Noda_pccp_16, alter_Naka_natcom_19, alter_Hayami_jpsj_19, alter_Yuan_prb_20}. In the altermagnet, there typically exist inequivalent nonmagnetic sites around a spin. Due to the inequivalence, a macroscopic time-reversal symmetry (mTRS) consisting of spin flip and lattice translation/inversion is broken under antiferromagnetic structures, resulting in spin splitting of the electron and magnon bands \cite{alter_Smejkal_prx_22_overview, alter_Noda_pccp_16, alter_Naka_natcom_19,  alter_Hayami_jpsj_19, alter_Yuan_prb_20, alter_Hernandez_prl_21, alter_Smejkal_prx_22, alter_Yuan_prb_21, alter_naka_npjSpin_25}. Since the spin splitting occurs even in the absence of both a spin-orbit coupling and a net magnetization, it is expected to serve for distinct types of spin current generator. A similar band splitting also emerges in compensated ferrimagnets where mutually compensating different magnetic sites inherently possess an inequivalence and thus, the mTRS is broken \cite{cferri_Groot_prl_95, cferri_Pickett, alter_Mazin_review_24,cferri_Liu_prl_25}.  

The mTRS breaking$-$the key factor behind the spin splitting$-$can arise not only from the crystallographic site inequivalence but also from different kinds of inequivalence; it has recently been proposed that orbital ordering on a regular lattice \cite{orbital_alter_Mook_prl_24}, spin-orbital-coupled ordering on an amorphous lattice \cite{spin-orbital_Ornellas_prb_26}, and spin cluster formation \cite{cluster_altermag} can yield altermagnetism and that a molecular dimer formation leads to compensated ferrimagnetism \cite{cferri_Misawa_prl_24}. Against this background, we previously showed that multiple-${\bf Q}$ nature of an antiferromagnetic structure can also drive the mTRS breaking. A triple-${\bf Q}$ collinear state emerging in a $J_3$-dominant kagome antiferromagnet possesses a spatial inhomogeneity, i.e., a kind of inequivalence, in its magnetic pattern, which leads to the mTRS breaking and the spin splitting of compensated-ferrimagnetic type \cite{cferri_KA_npjS_26}. 
In this paper, we extend the theory, and investigate the influence of an external magnetic field on the emergent triple-${\bf Q}$ compensated-ferrimagnet state, also analyzing the altermagnet for reference. 
More specifically, we discuss the field effects on the band structure and associated transport properties of the spin-split magnons in both the altermagnet and the emergent compensated ferrimagnet, bearing insulating systems in our mind.

\begin{figure}[t]
\begin{center}
\includegraphics[width=\columnwidth]{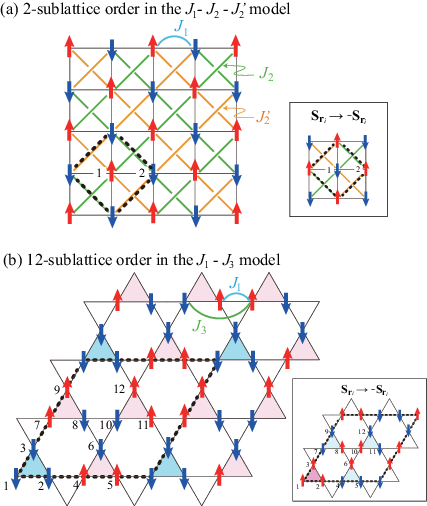}
\caption{Schematic pictures of the models and their collinear antiferromagnetic ground states at zero field, where the magnetic unit cells are indicated by dotted lines. (a) The 2-sublattice altermagnetic state in the $J_1$-$J_2$-$J_2^\prime$ model on the square lattice, where cyan, green, and orange lines represent the exchange interactions $J_1$, $J_2$, and $J_2^\prime$, respectively. (b) The 12-sublattice state with compensated ferrimagnetic nature in the $J_1$-$J_3$ model on the kagome lattice, where cyan and green curves represent $J_1$ and $J_3$, respectively. In (b), pink and cyan upward triangles possess the local magnetization of $+1$ (two-up and one-down) and $-3$ (three-down), respectively. In each of (a) and (b), numbers in the magnetic unit cell denote sublattice indices, and the inset shows the spin-flipped configuration. A cross at the center of the hexagon with $\uparrow$ spins on its six corners indicates a rotational center. In the text, the original (main panel) and spin-flipped (inset) configurations possessing $-3$ and $+3$ triangles are called $-$ and $+$ states, respectively. \label{fig1} }
\end{center}
\end{figure}

A minimal model for the altermagnet would be the $J_1$-$J_2$-$J_2^\prime$ model on the square lattice shown in Fig. \ref{fig1} (a) \cite{alter_model_Brekke_prb_23, alter_model_Cichutek_prr_25, alter_model_Eto_prb_25}. For antiferromagnetic $J_1>0$ and ferromagnetic $J_2, J_2^\prime <0$, the conventional 2-sublattice collinear antiferromagnetic structure is realized at zero field. Due to the inequivalence between $J_2$ and $J_2^\prime$, which originates from a crystallographic asymmetry of the nonmagnetic sites around the spin, the spin-flipped state [see the inset of Fig. \ref{fig1} (a)] cannot be superimposed onto the original configuration by the lattice translation, so that the mTRS is broken and resultantly, the spin splitting occurs. Since the flipped and original states can be superimposed by the $\frac{\pi}{2}$ rotation, the up- and down-spin magnon bands are interconnected by this rotation, showing the $d$-wave-type spin alternation in the momentum space [see the right panel of Fig. \ref{fig3} (a)]. 

In contrast to the altermagnet, the emergent compensated ferrimagnet is realized in the conventional spin model with symmetric exchange interactions, the $J_1$-$J_3$ model on the kagome lattice shown in Fig. \ref{fig1} (b). Dominant antiferromagnetic $J_3>0$ triggers ordering of a commensurate triple-${\bf Q}$ state with a 12-sublattice collinear structure which consists of three $\uparrow\uparrow\downarrow\downarrow$ chains running along the three bond directions \cite{RMO-collinear_Grison_prb_20, KagomeSkX_AK_22, Oct-collinear_Kiese_prr_23}.  
As indicated by cyan and pink colored triangles in Fig. \ref{fig1} (b), this antiferromagnetic structure contains two types of upward triangles within the magnetic unit cell (dotted parallelogram), the three-down triangle and the two-up and one-down triangle, each having a plaquette magnetization of ${\bf m}_\triangle=-3$ (cyan) and +1 (pink), respectively, where ${\bf m}_\triangle$ is defined as ${\bf m}_\triangle ={\bf S}_A +{\bf S}_B+{\bf S}_C$ with three corner spins of the triangle ${\bf S}_A$, ${\bf S}_B$, and ${\bf S}_C$. Since these inequivalent ${\bf m}_\triangle$'s are cancelled out within the unit cell, the state can be viewed as a compensated ferrimagnet in units of the triangle plaquette  \cite{cferri_KA_npjS_26}. 
Since the spin-flipped state contains the three-up triangle [see the inset of Fig. \ref{fig1} (b)] which does not exist in the original state, the flipped and original configurations cannot be superimposed by any spatial symmetry operations. The broken mTRS of this kind leads to the $s$-wave spin splitting [see the right panel of Fig. \ref{fig7} (a) below]. Throughout this paper, the former (latter) spin configuration with ${\bf m}_\triangle=-3$ (${\bf m}_\triangle=+3$) will be called a $-$ (+) state. These $+$ and $-$ states do not possess net magnetization and are degenerate at $H=0$.
 
One of the distinguished functionalities of these spin-split magnons would be spin-current generation at zero field. Although in the conventional antiferromagnet, the spin current cannot be generated by the thermal gradient at zero field as up- and down-spin magnon flows are canceled out, in the spin-split magnets, the thermal activation of the spin current is possible as the up- and down-spin degeneracy is lifted \cite{alter_Naka_natcom_19, cferi_Seebeck_Maekawa_prb_13, cferri_Lee_arX_26, cferri_KA_npjS_26}. Then, the question is how an external magnetic field affects the up- and down-spin magnon bands and resultantly, magnon transports. The application of the magnetic field, on the other hand, often leads to spin canting, destabilizing the collinear spin configuration.  
In this paper, we introduce an easy-axis anisotropy to make the collinear state robust against the weak parallel magnetic field, and examine the field effect on the spin-split magnon and its transport in the easy-axis collinear state.

We show that for positive (negative) magnetic field $H$ whose energy scale is smaller than the anisotropy gap, the up- and down-spin magnon bands are shifted to lower (higher) and higher (lower) energies, respectively, similarly to the case of the Zeeman coupling in electron systems. Thus, for $H>0$, the lower-energy bands tend to be positively spin polarized. Resultantly, in the altermagnet, the $d$-wave spin alternation tends to be deformed into the $s$-wave one, which is reflected as the change in the direction of the spin current generated by thermal gradient. In the compensated ferrimagnet, the $s$-wave spin splitting at $H=0$, i.e., the population imbalance between the up- and down-spin magnons at $H=0$, can induce an asymmetric field dependence of the longitudinal spin and thermal conductivities when the magnetic state remains unchanged during a field sweep.

The outline of this paper is as follows: in Sec. II, we introduce the models and explain how to calculate the magnon band and transport coefficients. Results for the altermagnet are discussed in Sec. III, and this is followed by Sec. IV in which we show results for the emergent compensated ferrimagnet. We end the paper with summary and discussion in Sec. V, where experimental implications for altermagnets such as $\alpha$-Fe$_2$O$_3$ are addressed. Details of analytical calculations and Monte Carlo (MC) simulations are provided in Appendices A and B, respectively, where the latter is used to determine temperature and magnetic-field phase diagrams shown in Figs. \ref{fig2} and \ref{fig6}. 

\section{Theoretical framework}
In this section, we first introduce two models; one is for the altermagnet and the other is for the compensated ferrimagnet. We then explain the method to calculate the magnon band and the transport coefficients, which is common to the two models once the spin Hamiltonians are rewritten into the Bogoliubov magnon Hamiltonians.    
\subsection{Model}
Classical spin models with the easy-axis anisotropy $D$ and the magnetic field along the easy axis $H$ are our starting point. Throughout this paper, ${\bf S}_{{\bf r}_i}$ represents a classical Heisenberg spin at lattice site ${\bf r}_i$, and the lattice constant is taken to be 1.

As explained in Sec. I, a fundamental aspect of the altermagnet can be captured by the $J_1$-$J_2$-$J_2^\prime$ model on the square lattice \cite{alter_model_Brekke_prb_23, alter_model_Cichutek_prr_25, alter_model_Eto_prb_25} shown in Fig. \ref{fig1} (a). Since the second NN interactions $J_2$ and $J_2^\prime$ are different from each other, the crystal unit cell contains two inequivalent sites which are denoted by 1 and 2 in Fig. \ref{fig1} (a). Then, the spin Hamiltonian is given by
\begin{eqnarray}\label{eq:Hamiltonian_s_alter}
{\cal H}_{\rm alt}  &=& \sum_{i\in 1, 2} \frac{J_1}{2} \sum_{l=1}^4{\bf S}_{{\bf r}_i}\cdot {\bf S}_{{\bf r}_i +{\bf a}_l} \nonumber\\
&+& \sum_{i\in 1}\Big(J_2 \sum_{l=1}^2 {\bf S}_{{\bf r}_i}\cdot {\bf S}_{{\bf r}_i +{\bf d}_l}  + J_2^\prime \sum_{l=3}^4 {\bf S}_{{\bf r}_i}\cdot {\bf S}_{{\bf r}_i +{\bf d}_l} \Big) \nonumber\\
&+& \sum_{i\in 2}\Big(J_2^\prime \sum_{l=1}^2 {\bf S}_{{\bf r}_i}\cdot {\bf S}_{{\bf r}_i +{\bf d}_l}  + J_2 \sum_{l=3}^4 {\bf S}_{{\bf r}_i}\cdot {\bf S}_{{\bf r}_i +{\bf d}_l} \Big) \nonumber\\
& - & D \sum_{i\in 1,2} (S_{{\bf r}_i}^z)^2 - H \sum_{i\in 1,2} S_{{\bf r}_i}^z ,
\end{eqnarray}
where $J_1$ and $J_2\neq J_2^\prime$ are the NN antiferromagnetic and the second NN ferromagnetic interactions, respectively, and the associated real-space vectors are defined by ${\bf a}_1=-{\bf a}_2=(1,0)$, ${\bf a}_3=-{\bf a}_4=(0,1)$, ${\bf d}_1=-{\bf d}_2=(1,1)$, and ${\bf d}_3=-{\bf d}_4=(-1,1)$. In the case of $D=0$ and $H=0$, the ground state is the 2-sublattice collinear antiferromagnetic order, and the magnetic unit cell indicated by the dotted parallelogram in Fig. \ref{fig1} (a) is the same as the crystallographic one. Due to the asymmetry in the interactions $J_2$ and $J_2^\prime$, the altermagnetic spin-splitting occurs \cite{alter_model_Brekke_prb_23, alter_model_Cichutek_prr_25, alter_model_Eto_prb_25}. 

 As for the compensated ferrimagnet, we introduce the $J_1$-$J_3$ model on the kagome lattice \cite{cferri_KA_npjS_26} shown in Fig. \ref{fig1} (b). Since the NN and third NN interactions $J_1$ and $J_3$ are symmetric around each spin, the spin Hamiltonian can simply be written as 
\begin{eqnarray}\label{eq:Hamiltonian_s_cfr}
{\cal H}_{\rm cfr}  &=& J_1 \sum_{\langle i,j \rangle } {\bf S}_{{\bf r}_i}\cdot{\bf S}_{{\bf r}_j} + J_3 \sum_{\langle \langle i,j \rangle \rangle} {\bf S}_{{\bf r}_i}\cdot{\bf S}_{{\bf r}_j}  \nonumber\\
&-& D \sum_i (S_{{\bf r}_i}^z)^2 - H \sum_i S_{{\bf r}_i}^z,
\end{eqnarray}
where $\langle \rangle$ and $\langle \langle  \rangle \rangle$ represent the summations over the NN and the third NN pairs, respectively. In the model, $J_3$ is active only along the bond direction, although in general, it can act on the off-bond third NN pairs. The $J_3$-dominant situation of this kind is reported to be realized in BaCu$_3$V$_2$O$_8$(OH)$_2$ \cite{CoplanarOct_Boldrin_prl_18}. It is known that for relatively strong antiferromagnetic $J_3$ roughly satisfying $J_3>|J_1|$, the 12-sublattice collinear antiferromagnetic state is stabilized at $H=0$ \cite{RMO-collinear_Grison_prb_20, KagomeSkX_AK_22}, where the state is characterized by the three commensurate ordering vectors ${\bf Q}_1=\frac{\pi}{2}(-1,-\frac{1}{\sqrt{3}})$, ${\bf Q}_2=\frac{\pi}{2}(1,-\frac{1}{\sqrt{3}})$, and ${\bf Q}_3=\frac{\pi}{2}(0,\frac{2}{\sqrt{3}})$. In Fig. \ref{fig1} (b), the 12 sublattices within the magnetic unit cell (dotted parallelogram) are labeled by 1-12. Due to the emergent inhomogeneous pattern of the plaquette magnetization ${\bf m}_\triangle$, the compensated-ferrimagnetic spin-splitting occurs \cite{cferri_KA_npjS_26}.

In the presence of the easy-axis anisotropy $D>0$, the collinear axis merely orients in the easy-axis direction, and thus, it turns out that the antiferromagnetic long-range orders occur at finite temperatures with the spin polarization axis along the easy-axis direction \cite{KagomeSkX_AK_23}. In the magnetic field $H$ applied parallel to the easy axis, the easy-axis anisotropy prevents the spin canting, so that the collinear antiferromagnetic orders remain robust as long as $H$ is small (see Figs. \ref{fig2} and \ref{fig6} and discussion below). 
 
 \subsection{Magnon band}    
In the spin models (\ref{eq:Hamiltonian_s_alter}) and (\ref{eq:Hamiltonian_s_cfr}), the ground states for small $H$ are the 2-sublattice and 12-sublattice collinear antiferromagnetic orders, respectively. In general, magnon excitations from $N$-sublattice collinear ground state can be examined by using the Holstein-Primakoff transformation which, for a site $i$ belonging to sublattice $\mu$ ($\mu=1$-$N$), is given by
\begin{equation}\label{eq:HP}
\left\{\begin{array}{l}
S^x_{{\bf r}_i} \simeq \sqrt{\frac{S}{2}} (\hat{a}_{{\bf r}_i}^\mu+\hat{a}_{{\bf r}_i}^{\mu\dagger}) \\
S^y_{{\bf r}_i} \simeq m_\mu \frac{1}{i}\sqrt{\frac{S}{2}} (\hat{a}_{{\bf r}_i}^\mu-\hat{a}_{{\bf r}_i}^{\mu\dagger}) \\
S^z_{{\bf r}_i}= m_\mu ( S- \hat{a}_{{\bf r}_i}^{\mu\dagger} \hat{a}_{{\bf r}_i}^\mu )
\end{array}\right. ,
\end{equation}
where $S$ is the size of the spin, $m_\mu=\pm 1$ represents the spin direction of the sublattice $\mu$, i.e., $\uparrow$ or $\downarrow$, and $\hat{a}_{{\bf r}_i}^{\mu\dagger}$ ($\hat{a}_{{\bf r}_i}^\mu$) is the creation (annihilation) operator of the magnon. 
By introducing the basis vector $\mbox{\boldmath $\Phi$}_{B{\bf q}}^\dagger =(\hat{a}_{\bf q}^{1\, \dagger},  \hat{a}_{\bf q}^{2\, \dagger}, \cdots , \hat{a}_{\bf q}^{N \, \dagger},  \hat{a}_{-{\bf q}}^{1}, \hat{a}_{-{\bf q}}^{2}, \cdots , \hat{a}_{-{\bf q}}^{N} )$ with the Fourier transforms of the magnon operators $\hat{a}_{\bf q}^{\mu \dagger}$ and $\hat{a}_{\bf q}^{\mu}$, one can rewrite an original spin Hamiltonian $\cal{H}_S$ into the quadratic form as
\begin{equation}\label{eq:Hamiltonian_magnon}
 {\cal H}_S\simeq \frac{S}{2} \sum_{\bf q} \mbox{\boldmath $\Phi$}_{B{\bf q}}^\dagger  H_{B{\bf q}} \mbox{\boldmath $\Phi$}_{B{\bf q}}, \quad H_{B{\bf q}}= \left(\begin{array}{cc}
A_{\bf q} & B_{\bf q}  \\
B_{{\bf q}} & A_{{\bf q}} 
\end{array} \right),
\end{equation}
where a constant contribution has been omitted. The concrete expressions of $N\times N$ matrices $A_{\bf q}$ and $B_{\bf q}$ in the two cases of Eqs. (\ref{eq:Hamiltonian_s_alter}) and (\ref{eq:Hamiltonian_s_cfr}) are given in Appendix A.  

The magnon Hamiltonian (\ref{eq:Hamiltonian_magnon}) can be diagonalized with the use of a para-unitary matrix $T_{\bf q}$ as \cite{LinearRes_Matsumoto_prb_14} 
\begin{equation}\label{eq:eigen}
T_{\bf q}^\dagger H_{B{\bf q}} T_{\bf q}=\left( \begin{array}{cc}
E_{\bf q} & 0 \\
0 & E_{-{\bf q}} 
\end{array} \right) , 
E_{\bf q} = \left(\begin{array}{ccc}
\epsilon_{1, {\bf q}} & & 0 \\
 & \ddots & \\
 0 && \epsilon_{N, {\bf q}} 
\end{array} \right) . \nonumber
\end{equation}
Since $\sigma_3 H_{B{\bf q}} T_{\bf q}= T_{\bf q} \left( \begin{array}{cc}
E_{\bf q} & 0 \\
0 & -E_{-{\bf q}} 
\end{array} \right)$ with $\sigma_3 = I_{N\times N} \otimes (-I)_{N\times N}$ is satisfied, $\epsilon_{n, {\bf q}}$ and $T_{\bf q}$ can be obtained by solving the eigen value problem for $\sigma_3 H_{\bf q}$. 

To characterize the magnon band $\epsilon_{n, {\bf q}}$, we introduce the magnon ''spin'' $h_{n,{\bf q}} = \sum_{\mu=1}^{N} (-m_\mu) ( |T_{{\bf q},\mu n}|^2 + |T_{{\bf q},\mu n+N}|^2)$ \cite{cferri_KA_npjS_26}. The meaning of this quantity $h_{n,{\bf q}}$ can be understood in the following way. As one can see from the expression of $S^z_{{\bf r}_i}$ in Eq. (\ref{eq:HP}), the excitation of $\hat{a}_{{\bf r}_i}^{\mu}$ magnon reduces the size of the local moment by $S^z_{{\bf r}_i} - m_\mu S = -m_\mu \hat{a}_{{\bf r}_i}^{\mu\dagger} \hat{a}_{{\bf r}_i}^\mu$ which corresponds to the spin polarization carried by the magnon when it moves. When a positive (negative) field $H$ is applied, the down-spin (up-spin ) sublattice with $m_\mu=-1$ ($m_\mu=+1$) becomes locally unstable, and resultantly, the oppositely polarized positive (negative) magnons can easily be excited. In the diagonalized $n$th magnon band $\epsilon_{n,{\bf q}}$, such sublattice $\hat{a}_{{\bf r}_i}^{\mu}$ magnons are hybridized, so that we shall evaluate the effective spin polarization of the $n$th magnon band from the hybridization contribution. Noting that the net fluctuation contribution is given by
$\sum_{\bf q}\sum_{\mu=1}^{N}(-m_\mu\langle \hat{a}^{\mu \dagger}_{\bf q}\hat{a}^{\mu}_{\bf q} \rangle)=\sum_{\bf q}\sum_{n =1}^{N} f_B(\epsilon_{n,{\bf q}}) h_{n,{\bf q}} + \delta M_0$ with the Bose distribution function $f_B(x)=\frac{1}{e^{x/T}-1}$ and the zero-point fluctuation contribution $\delta M_0 = \sum_{\bf q} \sum_{n,\mu=1}^N (-m_\mu)  |T_{{\bf q},\mu n+N}|^2$, it turns out that the prefactor of the Bose distribution function $h_{n,{\bf q}}$ can be interpreted as the effective spin polarization of the $n$th magnon band. As discussed in Ref. \cite{cferri_KA_npjS_26}, $h_{n,{\bf q}}$ can also be interpreted as a generalized form of the magnon chirality \cite{alter_magnon_Smejkal_prl_23, alter_magnon_Liu_prl_24,MagnonPol_Nambu_prl_20, MagnonPol_Kawamoto_apl_24}. 

As we will see in Secs. III and IV, similarly to the Zeeman coupling in electron systems, a positive (negative) magnetic field of $H>0$ ($H<0$) lowers the energy of positively (negatively) polarized magnon branches of $h_{n,{\bf q}}>0$, while it raises the opposite-spin branches of $h_{n,{\bf q}}<0$.

\subsection{Magnon transport}
In this paper, we investigate the spin and thermal currents carried by the spin-split magnons under a thermal gradient applied in the two-dimensional lattice plane [see the insets of Figs. \ref{fig4} (b) and \ref{fig8} (b)]. The longitudinal spin response in this setup corresponds to the lateral spin Seebeck coefficient. 

For the spin Hamiltonian ${\cal H}_{S}= \sum_{i,j}J_{ij} {\bf S}_{{\bf r}_i}\cdot {\bf S}_{{\bf r}_j} = \sum_i {\bf S}_{{\bf r}_i} \cdot \big( \sum_j J_{ij} {\bf S}_{{\bf r}_j} \big) \equiv \sum_i H({\bf r}_i)$, the spin and thermal currents can be derived from the conservation laws for the easy-axis $z$ component of the magnetization and the total energy as ${\bf J}_s^z = \sum_i {\bf r}_i \, \frac{d}{dt}S^z_i = i \sum_i {\bf r}_i \, [{\cal H}_S, S^z_i]$ and ${\bf J}_{th} = \sum_i {\bf r}_i \, \frac{d}{dt} H({\bf r}_i) = i \sum_i {\bf r}_i \, [{\cal H}_S, H({\bf r}_i)]$, respectively \cite{book_Mahan}. In the spin representation, these quantities are related to spin chiralities \cite{SpinDyn_Kawasaki_68, Thermal_Huber_ptp_68,  SpinDyn_Zotos_prb_05, SpinDyn_Jencic_prb_15, MHall_Mook_prb_16, MHall_Mook_prb_17, trans-sq_AK_prb_19, trans-tri_AK_prl_20, trans-cubic_A_prb_22}, whereas in the magnon representation, they are expressed as $J^z_{s,\mu}=\sum_{\bf q} \mbox{\boldmath $\Phi$}_{B{\bf q}}^\dagger  V^s_{{\bf q},\mu} \mbox{\boldmath $\Phi$}_{B{\bf q}} $ with $V^s_{{\bf q},\mu}= \frac{S}{4} (v_{{\bf q},\mu} I_m + I_m v_{{\bf q},\mu})$ and $J_{th,\mu}=  \sum_{\bf q} \mbox{\boldmath $\Phi$}_{B{\bf q}}^\dagger V^{th}_{{\bf q},\mu}  \mbox{\boldmath$\Phi$}_{B{\bf q}}$ with $V^{th}_{{\bf q},\mu} = \frac{S^2}{4}( v_{{\bf q},\mu} \sigma_3 H_{B{\bf q}}+ H_{B{\bf q}} \sigma_3 v_{{\bf q},\mu}) $ \cite{cferri_KA_npjS_26, LinearRes_Matsumoto_prb_14}. Here, $v_{{\bf q},\mu}$ is the velocity matrix $v_{{\bf q},\mu} = \frac{\partial}{\partial q_\mu} H_{B{\bf q}}$, and $I_m$ is the diagonal matrix consisting of $m_i=\pm 1$, the sign of the $i$th-sublattice spin polarization, which is given by $I_m= (\sigma_3)_{ij} (-m_j)$. For the magnon spin current ${\bf J}_s^z$ and the magnon thermal current ${\bf J}_{th}$, we calculate the transport coefficients in the linear response theory.

The spin current generated by the thermal gradient is characterized by the associated conductivity $\chi^{\rm SC}_{\mu \nu} = \frac{i}{T} \frac{d Q_{\mu\nu}(\omega + i 0 )}{d \omega}|_{\omega=0}$ with $Q_{\mu \nu}(i\omega_m) = - \frac{1}{V}\int_0^{1/T}\langle T_\tau J^z_{s,\mu} (\tau) J_{th, \nu}(0) \rangle e^{i\omega_m \tau} d\tau$. In general, the magnon transport depends on details of the magnon damping $\gamma$ originating from, for example, magnon-magnon and magnon-phonon scatterings \cite{Magdamp_Cornelissen_prb_16}. Nevertheless, as our interest here is in qualitative understanding of how the structure of the spin-split magnon band is reflected in the transport, we use, for simplicity, a constant-$\gamma$ approximation, or equivalently, a relaxation-time approximation, ignoring the wave number, energy, and temperature dependences of $\gamma$. Within this approximation, we can evaluate $\chi^{\rm SC}_{\mu\nu}$ as \cite{cferri_KA_npjS_26}
\begin{equation}\label{eq:response}
\chi^{\rm SC}_{\mu \nu} =  -\frac{1}{VT}\frac{1}{\gamma} \sum_{\bf q} \sum_{n=1}^{2N}  f^\prime_B(\epsilon_{n,{\bf q}})(T_{\bf q}^\dagger V^s_{{\bf q},\mu} T_{\bf q} )_{nn} (T_{\bf q}^\dagger V^{th}_{{\bf q},\nu}T_{\bf q})_{nn}  
\end{equation}
with $f^\prime_B(x) = \frac{d}{dx}f_B(x)$. In the same manner, the thermal conductivity $\kappa_{\mu \nu} $ can be calculated by simply replacing $V^s_{{\bf k},\mu}$ with $V^{th}_{{\bf k},\mu}$ in Eq. (\ref{eq:response}), and we have
\begin{equation}
\kappa_{\mu \nu} =  -\frac{1}{VT}\frac{1}{\gamma} \sum_{\bf q} \sum_{n=1}^{2N}  f^\prime_B(\epsilon_{n,{\bf q}})(T_{\bf q}^\dagger V^{th}_{{\bf q},\mu} T_{\bf q} )_{nn} (T_{\bf q}^\dagger V^{th}_{{\bf q},\nu}T_{\bf q})_{nn} . \nonumber
\end{equation}
Although the thermal conductivity can in principle be modified from the bare coefficient $\kappa_{\mu\nu}$ by the coupling between the spin and thermal currents \cite{book_Mahan, MHall_Mook_prb_17}, a quantitative evaluation of such effects is beyond the scope of this work. In the following sections, we will discuss the field dependence of the magnon band and the associated conductivities. 

\begin{figure}[t]
\begin{center}
\includegraphics[width=\columnwidth]{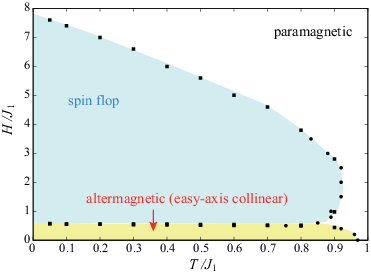}
\caption{Temperature and magnetic-field phase diagram of the easy-axis altermagnet on the square lattice with $J_2/J_1=-0.5$, $J_2^\prime /J_1 = -0.25$, and $D/J_1 = 0.02$. Yellow and cyan regions denote the stability regions of the easy-axis collinear and spin-flop states, respectively. The result is obtained in the MC simulation. \label{fig2} }
\end{center}
\end{figure}

\section{Field effects in the altermagnet}
In this section, we discuss the effects of the external magnetic field $H$ on the altermagnetic 2-sublattice collinear state realized in the $J_1$-$J_2$-$J_2^\prime$ model on the square lattice [Eq. (\ref{eq:Hamiltonian_s_alter})]. Since $J_2 \neq J_2^\prime$ is essential for the occurrence of the altermagnetic state, we use $J_2/J_1=-0.5$ and $J_2^\prime/J_1=-0.25$ as an example case. The easy-axis anisotropy $D$ is taken to be the small value of $D/J_1=0.02$.  

Figure \ref{fig2} shows the temperature and magnetic-field phase diagram obtained in the MC simulation (for details, see Appendix B). In the low-field region, the altermagnetic collinear state shown in Fig. \ref{fig1} (a) is stabilized without being subject to spin canting due to the easy-axis anisotropy. With increasing field strength, the spin-flop transition into the canted state occurs at $|H/J_1|=0.6$ which corresponds to the excitation gap stemming from the anisotropy. Hereafter, we will focus on the field range of $-0.6 \leq H/J_1 \leq 0.6$ in which the collinear state is robust. Although the sign of $H$ does not affect the static magnetic properties, i.e., the phase diagram, it turns out to affect dynamical properties of the system. 

\begin{figure}[t]
\begin{center}
\includegraphics[width=\columnwidth]{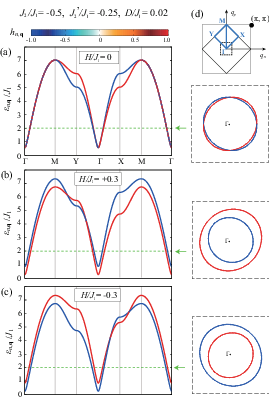}
\caption{Field dependence of the magnon band in the altermagnetic collinear state with the 2 sublattice structure (see the yellow region in Fig. \ref{fig2}). (a)-(c) The magnon dispersions at $H/J_1=0$, $+0.3$, and $-0.3$ (from top to bottom) along the path connecting high symmetry points shown in Fig. \ref{fig3} (d), where the color represents the spin polarization of the magnon $h_{n,{\bf q}}$ (reddish and bluish colors represent up and down spins, respectively). In each figure, the right panel shows the cross section of the magnon band at $\epsilon_{n, {\bf q}}/J_1 =2$, where the box enclosed by the dashed lines corresponds to the one near the $\Gamma$ point in (d). In (d), the inner diamond   denotes the Brillouin zone. The interaction parameters are the same as those in Fig. \ref{fig2}. \label{fig3} }
\end{center}
\end{figure}

\subsection{Magnon band}
Figure \ref{fig3} shows the field dependence of the magnon dispersion, where the color represents the spin of the $n$th magnon band $h_{n,{\bf q}}$. As reported elsewhere \cite{alter_model_Brekke_prb_23, alter_model_Cichutek_prr_25, alter_model_Eto_prb_25}, the magnon band exhibits a spin splitting at $H=0$, which can clearly be seen near the $X$ and $Y$ points [see Fig. \ref{fig3} (a) together with Fig. \ref{fig3} (d)]. Along the path connecting the $Y$ and $X$ points via the $\Gamma$ point, the splitting changes its sign, i.e., the lower energy branch changes from the negative $h_{n,{\bf q}}$ (down spin) state to the positive $h_{n,{\bf q}}$ (up spin) state. This sign change in $h_{n,{\bf q}}$ is reflected as a $d$-wave-type spin alternation in the contour plot of the band shown in the right panel of Fig. \ref{fig3} (a). 

As shown in Fig. \ref{fig3} (b), in the presence of the positive magnetic field $H>0$, the up-spin (down-spin) magnon band is shifted to the lower (higher) energy side and as a result, the splittings near the $X$ and $Y$ points are suppressed and enhanced, respectively. Such an anisotropic change in the splitting width manifests itself as a uniaxial deformation of the lowest energy band in the contour plot shown in the right panel. In the case of the negative magnetic field $H<0$, the opposite situation occurs: the up-spin (down-spin) magnon band is pushed up (down) and the spin splitting near the $X$ ($Y$) point is enhanced (suppressed) [see Fig. \ref{fig3} (c)]. Correspondingly, in the contour plot shown in the right panel of Fig. \ref{fig3} (c), the band deformation axis is rotated by 90$^\circ$. From these observations, we could say that the field-induced shift of the magnon band is quite similar to the Zeeman shift in electron systems. 

In the contour plots in Figs. \ref{fig3} (a)-(b), one notices that the $d$-wave-type spin alternation cannot be seen in the magnetic field, which could be understood in the following way. At $H=0$, the time reversal operation acts only on the spins, whereas at $H\neq 0$, it acts on both the spin and the magnetic field and thus, the time-reversed state cannot be superimposed onto the original one as the field direction remains flipped. This situation is similar to that of the compensated ferrimagnet in the sense that the original and time-reversed states cannot be interconnected by any spatial symmetries, which supports the occurrence of the $s$-wave nature of the spin splitting in the magnetic field.
  
\begin{figure}[t]
\begin{center}
\includegraphics[width=\columnwidth]{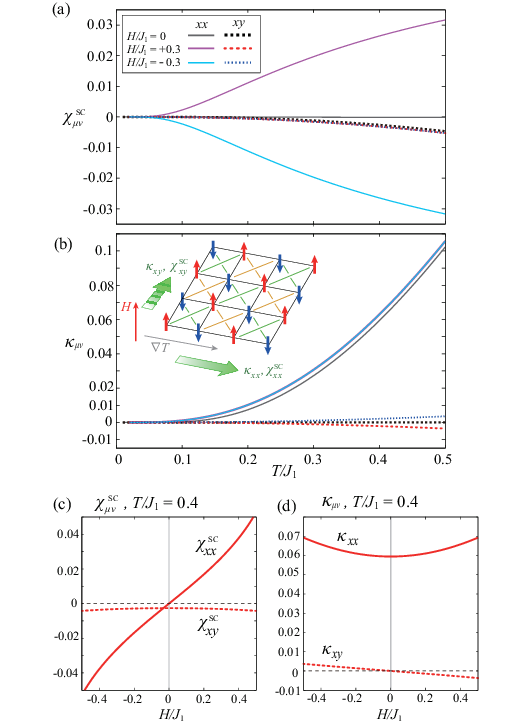}
\caption{Spin and thermal currents driven by thermal gradient in the altermagnetic state [see the system setup shown in the inset of (b)]. (a) Temperature dependence of the spin-current conductivity $\chi_{\mu\nu}^{\rm SC}$ and (b) that of the thermal conductivity $\kappa_{\mu\nu}$, where black, reddish, and bluish solid (dashed) curves represent to the longitudinal $xx$ (transverse $xy$) components at $H/J_1=0$, $+0.3$, and $-0.3$, respectively. Other parameters are the same as those in Figs. \ref{fig2} and \ref{fig3}. In (b), the $xx$ components for $H/J_1=\pm 0.3$ (magenta and cyan curves) exactly overlap. (c) and (d) Associated field dependences of (c) $\chi_{\mu\nu}^{\rm SC}$ and (d) $\kappa_{\mu\nu}$ at $T/J_1=0.4$. 
$\chi^{\rm SC}_{\mu\nu}$ and $\kappa_{\mu\nu}$ are normalized by $S^3J_1/\gamma$ and $S^4J_1^2/\gamma$, respectively.\label{fig4} }
\end{center}
\end{figure}

\subsection{Magnon transport}
As discussed above, the magnon band responds to the positive and negative magnetic fields in different manners, so that the associated magnon transport should also behave differently.   
Figures \ref{fig4} (a) and (b) show the temperature dependences of the spin-current conductivity $\chi_{\mu\nu}^{\rm SC}$ and the thermal conductivity $\kappa_{\mu\nu}$ at $H/J_1=0$ (black symbols), $H/J_1=+0.3$ (reddish symbols), and $H/J_1=-0.3$ (bluish symbols) for the system setup shown in the inset of Fig. \ref{fig4} (b) where the temperature gradient is applied in the lateral direction. Since the $yy$ ($xy$) components of $\chi_{\mu\nu}^{\rm SC}$ and $\kappa_{\mu\nu}$ are identical to the associated $xx$ ($xy$) components at $H=0$ \cite{alter_Naka_natcom_19, Alt_magtrans_Wu_cpl_25} and $H\neq 0$ as well, only the $xx$ and $xy$ components are shown in Fig. \ref{fig4}. Note that $\chi_{xy}^{\rm SC}=\chi_{yx}^{\rm SC}\neq 0$ or $\kappa_{xy}=\kappa_{yx}\neq 0$ does not mean the occurrence of the Hall effect because it is usually defined by the antisymmetric part of the transverse conductivity, i.e., $\chi_{xy}^{\rm SC}-\chi_{yx}^{\rm SC}$ or $\kappa_{xy}-\kappa_{yx}$. Since the symmetric conductivity tensor $\left(\begin{array}{cc}
\chi_{xx}^{\rm SC} & \chi_{xy}^{\rm SC} \\
\chi_{yx}^{\rm SC} & \chi_{yy}^{\rm SC} \end{array}\right)$ with $\chi_{xy}^{\rm SC}=\chi_{yx}^{\rm SC}\neq 0$ can be diagonalized by a rotation matrix, the existence of the symmetric off-diagonal components means that principal axes are tilted from the $x$ and $y$ axes.

\begin{figure}[t]
\begin{center}
\includegraphics[width=\columnwidth]{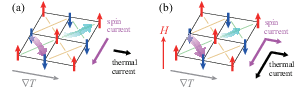}
\caption{A schematic image of the origin of the longitudinal and transverse responses in the altermagnetic states at (a) $H/J_1=0$ and (b) $H/J_1=+0.3$, where magenta and cyan arrows represent the flow of the positively and negatively polarized magnons, respectively. \label{fig5} }
\end{center}
\end{figure}

One can see from Fig. \ref{fig4} (a) that at $H=0$, the spin conductivity $\chi_{\mu\nu}^{\rm SC}$ is nonzero only for the transverse $xy$ component, while the thermal conductivity $\kappa_{\mu\nu}$ is nonzero only for the longitudinal $xx$ component, as reported in a similar altermagnetic state \cite{alter_Naka_natcom_19}. This is the magnon analogue to the spin-splitter effect in the metallic altermagnet \cite{alter_Hernandez_prl_21}, and can be understood from the magnon band structure shown in the right panel of Fig. \ref{fig3} (a). 
The short principal axis of the ellipsoid magnon band with up spin (down spin) is tilted from the $q_x$ axis by $-45^\circ$ ($+45^\circ$), so that the up-spin (down-spin) magnon flow biased by the thermal gradient in the $+x$ direction is accompanied by the $-y$ ($+y$) component. Thus, the spin current, i.e., a difference between the up-spin and down-spin magnon flows, appears in the $-y$ direction, whereas the thermal current, i.e., the net magnon flow, appears in the $+x$ direction, being consistent with the calculation result of $\chi_{\mu\nu}^{\rm SC}$ and $\kappa_{\mu\nu}$.
The explanation based on the band structure can be understood more intuitively from the real space structure, as shown in Fig. \ref{fig5} (a). 

Suppose that the temperature gradient is applied in the $+x$ direction. Noting that the spin length is modified as $S-\sum_{\bf q}\hat{a}^{\mu\dagger}_{\bf q} \hat{a}^\mu_{\bf q}$, the magnon excitation at the down-spin (up-spin) sublattice turns out to yield the positively (negatively) polarized spin current, as indicated by the magenta (cyan) arrow in Fig. \ref{fig5} (a). Here, it is essential that due to the inequivalent second NN interactions $|J_2|>|J_2^\prime|$, the spin current flows (the magnon hops) dominantly in the $J_2$ direction. Since the $x$ ($y$) components of the positively and negatively polarized spin currents are parallel (antiparallel) to each other, the net spin and thermal currents flow along the $y$ and $x$ directions, respectively, at $H=0$ where the populations of the up- and down-spin magnons are the same.

In the presence of the positive magnetic field ($H>0$), the population of the up-spin magnon becomes larger than that of the down-spin one [see the right panel of Fig. \ref{fig3} (b)]. Then, as shown in Fig. \ref{fig5} (b), the spin current cancellation along the $x$ direction and the thermal current cancellation along the $y$ direction become incomplete, leading to the emergence of the additional components $\chi_{xx}^{\rm SC}\neq 0$ and $\kappa_{xy}\neq 0$. When the magnetic field is flipped to the opposite direction, the roles of the up- and down-spin magnons are interexchanged; the population of the down-spin magnon becomes larger [see the right panel of Fig. \ref{fig3} (c)], and $\chi_{xx}^{\rm SC}$ and $\kappa_{xy}$ change their signs with other responses being unchanged, as inferred from the field-flipped state of Fig. \ref{fig5} (b). This can readily be seen in the field dependences of the conductivities shown in Figs. \ref{fig4} (c) and (d).  

\section{Field effects in the emergent compensated ferrimagnet}
\begin{figure}[t]
\begin{center}
\includegraphics[width=\columnwidth]{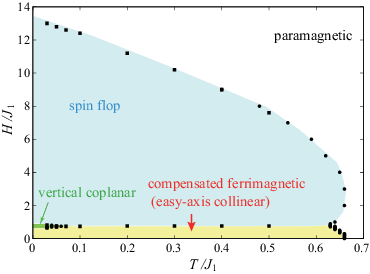}
\caption{Temperature and magnetic-field phase diagram of the easy-axis compensated ferrimagnet emerging on the kagome lattice with $J_3/J_1=1.2$ and $D/J_1 = 0.02$. Yellow, cyan, and green regions denote the stability regions of the easy-axis collinear, spin-flop, and vertical coplanar states, respectively. The result is obtained in the MC simulation. \label{fig6} }
\end{center}
\end{figure}

Now that the properties of the altermagnetic magnon are understood, we shall next discuss the field effects on the compensated ferrimagnetic state with the 12-sublattice collinear structure emerging in the $J_1$-$J_3$ model on the kagome lattice [Eq. (\ref{eq:Hamiltonian_s_cfr})].

\begin{figure}[t]
\begin{center}
\includegraphics[width=\columnwidth]{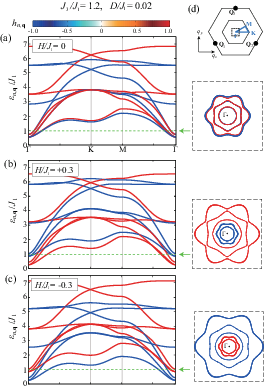}
\caption{Field dependence of the magnon band in the emergent compensated ferrimagnetic state with the 12-sublattice structure for $J_3/J_1=1.2$ and $D/J_1=0.02$ (see the yellow region in Fig. \ref{fig6}). (a)-(c) The magnon dispersions at $H/J_1=0$, $+0.3$, and $-0.3$ (from top to bottom) along the path connecting high symmetry points shown in Fig. \ref{fig7} (d), where notations are the same as those in Fig. \ref{fig3}. In (d), the inner small hexagon denotes the magnetic Brillouin zone. The result is obtained for the $-$ state with ${\bf m}_\triangle = -3$ shown in the main panel of Fig. \ref{fig1} (b). \label{fig7} }
\end{center}
\end{figure}

Figure \ref{fig6} shows the temperature and magnetic-field phase diagram obtained in the MC simulations for $J_3/J_1 =1.2$, the same parameter value as that in our previous work \cite{cferri_KA_npjS_26}, and $D/J_1=0.02$. One can see from Fig. \ref{fig6} that the compensated-ferrimagnetic state shown in Fig. \ref{fig1} (b) is robust against the field up to $H/J_1\sim 0.6$ at which the spin-flop transition occurs similarly to the case of the altermagnet shown in Fig. \ref{fig2}. 
The energy scale of $H/J_1\sim 0.6$ corresponds to the excitation gap originating from the easy-axis anisotropy [see Fig. \ref{fig7} (a)]. 
Near the spin-flop transition, a coplanar state with its plane being parallel to the field appears at low temperatures (for details, see Appendix B). 
As our interest is in the field effect on the easy-axis collinear state, we will hereafter focus on the low-field phase of Fig. \ref{fig6}.  

\subsection{Magnon band}
Figure \ref{fig7} shows the field dependence of the spin-split magnon band in the 12-sublattice compensated ferrimagnetic state shown in the main panel of Fig. \ref{fig1} (b), where positive and negative values of the magnon spin $h_{n,{\bf q}}$ are represented by reddish and bluish colors, respectively, as in Fig. \ref{fig3}. One can see from Fig. \ref{fig7} (a) that at $H=0$, the lowest and second-lowest energy bands are negatively and positively spin polarized, respectively, and these two do not intersect each other over the whole momentum space, exhibiting the $s$-wave type splitting in the contour plot shown in the right panel. In contrast to the altermagnet, the splitting in the present kagome system is determined not by the system parameters but by the magnetic structure. Since this magnetic structure remains stable over the large parameter space, the splitting  size normalized by the band width is also insensitive to the system parameters. By applying the positive (negative) magnetic field, positively (negatively) spin polarized bands are shifted to lower energies, and the opposite spin bands are shifted to higher energies, and as a result, the low energy excitations are dominated by the positive (negative) spin magnons [see the main and right panels in Figs. \ref{fig7} (b) and (c)]. Although in the present 12 sublattice state, the band structure is rather complicated, this Zeeman-shift-like behavior of the magnons is qualitatively the same as that for the altermagnet. In the compensated ferrimagnet, however, the spin polarization of the lowest energy magnon is changed from positive at $H=0$ to negative for $H<0$, whereas for $H>0$, it remains positive. 

In the case of Fig. \ref{fig7}, the lowest-energy magnon band at $H=0$ is negatively spin polarized, and this spin polarization distinguishes the positive and negative magnetic field. 
 What determines the spin polarization of the lowest energy band? The origin consists in the underlying magnetic structure shown in the main panel of Fig. \ref{fig1} (b) which contains the three-down triangle plaquette. Since the magnetic structure is selected as a result of the spontaneous symmetry breaking, the spin-flipped $+$ state with the three-up triangle plaquette shown in the inset of Fig. \ref{fig1} (b) can equally be the ground state. Which one of the two degenerate states is actually selected at $H=0$ depends on, for example, cooling processes or initial conditions. Once the spin-flipped $+$ state is stabilized instead of the $-$ state, the lowest-energy magnon band is positively spin polarized.
 In the next subsection, we will discuss how the feature distinguishing the positive and negative fields is reflected in the magnon transport. 

\begin{figure}[t]
\begin{center}
\includegraphics[width=\columnwidth]{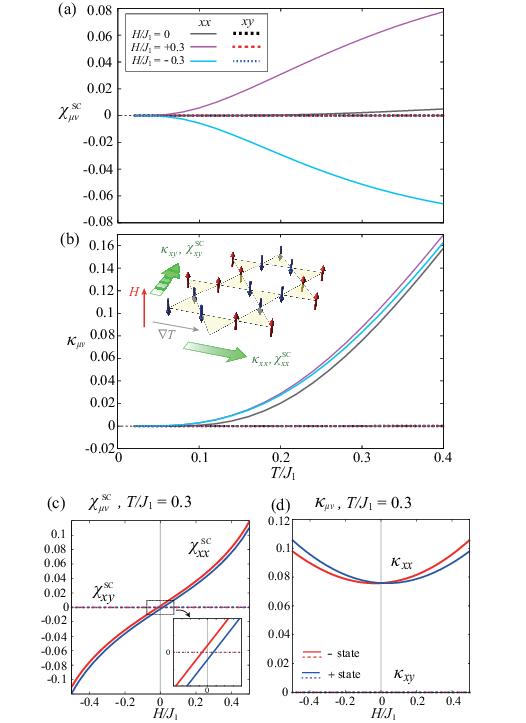}
\caption{Spin and thermal currents driven by thermal gradient in the $-$ state of the emergent compensated ferrimagnet for $J_3/J_1=1.2$ and $D/J_1=0.02$ [see the system setup shown in the inset of (b)]. (a) Temperature dependence of the spin-current conductivity $\chi_{\mu\nu}^{\rm SC}$ and (b) that of the thermal conductivity $\kappa_{\mu\nu}$, where the symbol notations are the same as those in Fig. \ref{fig4}. (c) and (d) Associated field dependences of (c) $\chi_{\mu\nu}^{\rm SC}$ and (d) $\kappa_{\mu\nu}$ at $T/J_1=0.3$, where the red and blue curves correspond to the $-$ and $+$ states, respectively. In (c), the inset shows the zoomed view of the region enclosed by a box in the main panel. Normalizations of $\chi^{\rm SC}_{\mu\nu}$ and $\kappa_{\mu\nu}$ are the same as those in Fig. \ref{fig4}. \label{fig8} }
\end{center}
\end{figure}

\subsection{Magnon transport}
Figures \ref{fig8} (a) and (b) show the temperature dependences of the spin-conductivity $\chi_{\mu\nu}^{\rm SC}$ and the thermal conductivity $\kappa_{\mu\nu}$ in the $-$ state of the emergent compensated ferrimagnet at $H/J_1=0$, +0.3, and -0.3. Since the $xx$ ($xy$) and $yy$ ($yx$) components of these conductivities are almost identical to each other, only the former is shown in Fig. \ref{fig8}. As one can see from the dashed curves in Fig. \ref{fig8}, the transverse components $\chi_{xy}^{\rm SC}$ and $\kappa_{xy}$ are always zero in the present case with the symmetric interactions, which is in contrast to the altermagnet where the asymmetric interaction yields the transverse responses. In other words, due to the isotropic $s$-wave nature of the spin splitting, the thermal gradient simply drives magnon flows in the thermal-gradient direction. 

The absence of the transverse response can also be understood from the viewpoint of the symmetry. As indicated from the 6-fold symmetric band structure, the spin configuration possesses the $D_{6h}$ point group symmetry \cite{cferri_KA_npjS_26}, where the rotational center is indicated by the cross in Fig. \ref{fig1} (b). Then, the conductivity tensor should also have the 6-fold symmetry, satisfying $\chi_{\mu\nu}^{\rm SC} = R_{\mu \rho} R_{\nu \eta} \chi_{\rho\eta}^{\rm SC}$ with $R=\left(\begin{array}{cc}
\cos \phi & -\sin \phi \\
\sin \phi & \cos \phi \end{array}\right)$ and $\phi=n\frac{2\pi}{6}$. From this equation, the relations $\chi_{xx}^{\rm SC} = \chi_{yy}^{\rm SC}$ and $\chi_{xy}^{\rm SC} = \chi_{yx}^{\rm SC}=0$ hold for the symmetric part of the conductivity tensor, and such a situation is also the case for the thermal conductivity $\kappa_{\mu \nu}$. As we will see below, although the transverse responses are absent, a feature characteristic of the compensated ferrimagnet shows up in the longitudinal conductivities $\chi_{xx}^{\rm SC}$ and $\kappa_{xx}$ under the magnetic field, which are represented by solid curves in Figs. \ref{fig8} (a) and (b). 

Before discussing the specific field dependence of $\chi_{xx}^{\rm SC}$ and $\kappa_{xx}$, we start from their behavior qualitatively expected from the band structure shown in Fig. \ref{fig7}. Since in the positive (negative) magnetic field, the lower-energy magnon bands are positively (negatively) polarized, it is expected that the thermally driven magnon flow, i.e., the longitudinal thermal current, is accompanied by the positive (negative) spin current. Although not only the lower-energy spin states but also higher-energy opposite-spin states contribute at moderate temperatures, the net spin current should be dominated by the lower-energy band as long as the energy difference between the up- and down-spin bands is large enough. Actually, as one can see from the in-field data (magenta and cyan solid curves) in Figs. \ref{fig8} (a) and (b), monotonic temperature dependences of $\chi_{xx}^{\rm SC}$ and $\kappa_{xx}$ are qualitatively the same, supporting the above consideration based on the band structure that the low-energy magnons carry both the thermal and spin currents.   

Although this behavior of the longitudinal conductivities is the same as that of the altermagnet [see Figs. \ref{fig3} (a) and (b)], a difference can appear in the field dependence of $\chi_{xx}^{\rm SC}$ and $\kappa_{xx}$. Figures \ref{fig8} (c) and (d) show the field dependences of $\chi_{\mu\nu}^{\rm SC}$ and $\kappa_{\mu\nu}$ at $T/J_1=0.3$ in the $-$ state (red curves). For comparison, the associated field dependences in the $+$ state are also shown (blue curves). One can see that $|\chi_{xx}^{\rm SC}|$ and $\kappa_{xx}$ for each state are asymmetric with respect to $H=0$. Such an unusual situation originates from the compensated ferrimagnetic nature; due to the $s$-wave spin splitting, the lowest energy band in the $-$ state is negatively polarized already at $H=0$ being independent of the wave number, and thus, the positive and negative fields act on the magnon band in different manners, leading to the asymmetric field dependence. It should be noted that in the compensated ferrimagnetic state, the spin current can thermally be driven in the longitudinal direction even at $H=0$ due to the spin splitting [see the inset of Fig. \ref{fig8} (c)], where the velocity difference between the up- and down-spin magnons is essential rather than the spin polarization of the lowest energy band itself as the energies of the lowest two up and down bands are close to each other \cite{cferri_KA_npjS_26}. 

In Figs. \ref{fig8} (c) and (d), there are two branches corresponding to the $\pm$ states. Then, one may wonder which one is actually selected. Also, the asymmetric field dependence of the transport coefficients at $T\neq 0$, $\kappa_{\mu\mu}(+H) \neq \kappa_{\mu\mu} (-H)$ and $|\chi_{\mu\mu}^{\rm SC} (+H)| \neq |\chi_{\mu\mu}^{\rm SC} (-H)|$ looks contradictory with the Onsager reciprocal relation. These issues can be understood as follows. Although the $H$ dependences represented by the red curves in Figs. \ref{fig8} (c) and (d) are obtained for the $-$ state being fixed, the inequality, .i.e., the asymmetric field dependence, suggests that the $–$ state is stable at $H>0$ while not at $H<0$ and the time-reversal counter part of the $+$ state can become favorable at $H<0$. Indeed, $\kappa_{\mu\mu}(+H)$ [$|\chi_{\mu\mu}^{\rm SC} (+H)|$] for the $-$ state and $\kappa_{\mu\mu}(-H)$ [$|\chi_{\mu \mu}^{\rm SC} (-H)|$] for the $+$ state turn out to be symmetric with respect to $H=0$ [compare the red and blue curves in Figs. 8 (c) and (d)]. Since at $T=0$, the $-$ and $+$ states are definitely degenerate as $H$ merely acts on their antiferromagnetic localized spins, the degeneracy lifting is due to thermal fluctuations. Actually, this tendency has been confirmed in field cooling runs of the Monte Carlo (MC) simulation. In the field sweep MC simulations at a fixed temperature, on the other hand, the state remains unchanged as is often the case with easy-axis magnets. Thus, in the field cooling process, the $-$ ($+$) state should be selected at $H>0$ ($H<0$) as the equilibrium state by the thermal fluctuation, and the resultant field dependence is symmetric due to the spin-flip on crossing $H=0$. In the field sweep process, on the other hand, the state remains without undergoing such a whole spin-flip and stays in the metastable state, leading to the asymmetric field dependences of $\kappa_{\mu\mu}(+H) \neq \kappa_{\mu\mu} (-H)$ and $|\chi_{\mu\mu}^{\rm SC} (+H) \neq \chi_{\mu \mu }^{\rm SC} (-H)|$.
 
 \section{Summary and Discussion}
In this paper, we theoretically investigate the easy-axis altermagnet and compensated ferrimagnet in a magnetic field, putting an emphasis on the field effects on the spin-split magnon band and the magnon transport under the thermal gradient. For the altermagnet and the compensated ferrimagnet, we analyze the 2-sublattice collinear state in the $J_1$-$J_2$-$J_2^\prime$ model on the square lattice and the triple-${\bf Q}$ 12-sublattice collinear state in the $J_1$-$J_3$ model on the kagome lattice, respectively. By changing the magnetic field from positive to negative with its strength weak enough such that the spin configuration remains collinear, we show that in both the spin-split systems, the up- and down-spin magnon bands are shifted to lower (higher) and higher (lower) energies by the positive (negative) magnetic field, respectively, similarly to the case of the Zeeman coupling in electron systems. In the, for example, positive field, the lower-energy magnon bands are positively spin polarized, so that the $d$-wave spin splitting of the altermagnet tends to be accompanied by the $s$-wave component characteristic of the compensated ferrimagnet. 
  
In the altermagnet, a transverse magnon spin current can be generated by the thermal gradient $\nabla T$ through the $d$-wave spin-splitting at $H=0$, and it becomes accompanied by a longitudinal component in the magnetic field where one of the up- and down-spin magnons is dominantly populated in the lower energy bands. In the compensated ferrimagnet with the $s$-wave spin splitting, the longitudinal spin current can thermally be activated even at $H=0$, as the lowest-energy magnon band is uniformly spin polarized over the whole momentum space. Reflecting the polarization selection at $H=0$, the field dependences of the spin and thermal conductivities $|\chi_{xx}^{\rm SC}|$ and $\kappa_{xx}$ can be asymmetric with respect to $H=0$ when the ordered state remains unchanged during the field sweep. These in-field properties are additional features of nonrelativistic spin splitting with various functionalities \cite{alter_Smejkal_prx_22_overview, alter_naka_npjSpin_25}.   

Our result above can briefly be summarized as follows: in the $d$-wave altermagnet with $\nabla T$ parallel to the spin-degenerate nodal direction (in the $s$-wave compensated ferrimagnet), the longitudinal and transverse spin-current responses are absent (present) and present (absent) at $H=0$, respectively, whereas at $H\neq 0$, the longitudinal component is induced (enhanced) with the transverse one remaining basically unchanged. The common feature of the two is the field-induced enhancement of the longitudinal response to the thermal gradient $\nabla T$. This originates from the Zeeman-like shift of the magnon bands; a positive (negative) magnetic field lowers the energy of positively (negatively) polarized magnon branches while it raises the opposite-spin branches, and resultantly, the positive (negative) magnons dominantly populated in the low energy states yield the longitudinal spin current when they are activated by the thermal gradient. Since the field-induced shift occurs in spin space being irrespective of the orbital structure of the splitting, the field-induced enhancement of the longitudinal spin-current response should commonly be observed in the spin-split collinear magnets including $g$-wave altermagnets.

As discussed in Ref. \cite{Magdamp_Cornelissen_prb_16}, the difference in the magnon scattering process is reflected in the temperature dependence of the damping coefficient and so would be true for the field dependence. Thus, specific functional forms of the temperature and field dependences of the transport coefficients, which are determined by the combination of the effects of the magnon band structure and damping, are material specific. Nevertheless, the qualitative aspect stemming  only from the band structure can directly be applied to the spin-polarized magnonic systems. The enhancement of the field-induced longitudinal response originates from the population imbalance between the up or down spin magnons, so that it should occur regardless of the damping effect.
 
We finally address experimental implications of our result. As our focus here is on magnon properties, relevant systems are nonmetallic magnets in which a conduction-electron spin current is basically absent. Several classes of altermagnet candidates are reported so far, among which $\alpha$-MnTe \cite{alter_magnon_Liu_prl_24} and $\alpha$-Fe$_2$O$_3$ \cite{alter_magnon_Sun_prl_25} are nonmetallic materials whose magnon band splittings are experimentally confirmed. In contrast to the simplified model of Eq. (\ref{eq:Hamiltonian_s_alter}) showing the $d$-wave splitting, these altermagnets exhibit $g$-wave type splittings due to the effect of further neighbor interactions. Nevertheless, the in-field properties discussed here are generic to the altermagnet with the easy-axis anisotropy, so that the qualitative description could be applied to $\alpha$-Fe$_2$O$_3$ possessing an easy-axis anisotropy at low temperatures ($\alpha$-MnTe is an easy-plane altermagnet); the Zeeman-like band shift should be induced by the magnetic field, and the spin and thermal currents driven by the thermal gradient should change their directions in the application of the magnetic field. Since these features should be observed as long as the energy scale of the magnetic field is smaller than the anisotropy gap, future experimental verification would be accessible. 
Actually, in hematite $\alpha$-Fe$_2$O$_3$, magnetic field effects have been studied in the context of the linear magnetostriction \cite{Hematite_LMS_Levitin_69, Hematite_LMS_BRomanov_93} and the spin current \cite{Hematite_SC_Lebrun_nature_18}. Although in the former, the applied field is usually tilted from the easy axis, in the latter, the spin-current measurements have been performed also for the parallel-field configuration as assumed here. With increasing parallel field, an injected spin current with its polarization parallel to the easy axis is gradually but non-monotonically suppressed toward the spin-flop transition \cite{Hematite_SC_Lebrun_nature_18}. Such a non-monotonic field dependence might be related to the altermagnetic nature, e.g., the field-induced directional change of the spin current as discussed in this paper. Although it would generally be a hard task to distinguish various effects possibly relevant to the spin transport, the combination of additional measurements, such as relaxometry with nitrogen-vacancy centers in diamond \cite{alter-NV_Bittencourt_prl_26}, could help elucidate the microscopic origin of the field-induced spin-transport phenomena.

Concerning the compensated ferrimagnet, fundamental properties of this class of magnets have extensively been studied \cite{cferri_review_jpsj_21}, where relevant materials usually have inequivalent magnetic sites with their spins compensating each other. A distinct feature of the present emergent compensated ferrimagnet is that all the magnetic sites are equivalent to one another and the ferrimagnetic cancellation occurs in units of the local magnetization on the triangle plaquette ${\bf m}_\triangle$; there are one ${\bf m}_\triangle =- 3$ (or $+3$) and three ${\bf m}_\triangle =+1$ (or $-1$) triangles within the magnetic unit cell. Since this superstructure is formed as a result of the spontaneous symmetry breaking, every plaquette has a chance to take ${\bf m}_\triangle = \pm 3$, suggesting that the system possesses configurational flexibility which could be controlled by temperature and magnetic fields. 
Such a flexibility might be a key factor in manipulating magnetic domains with ${\bf m}_\triangle \pm 3$.

The $J_1$-$J_3$ kagome antiferromagnet as a platform of the emergent compensated ferrimagnet, however, has not been reported so far in experiments. 
Although the vesignieite BaCu$_3$V$_2$O$_8$(OH)$_2$ is a $J_3$-dominant kagome antiferromagnet \cite{CoplanarOct_Boldrin_prl_18}, its 12 sublattice spin structure is not theoretically-expected collinear but rather coplanar. In the vesignieite, the unusual $J_3$-dominant situation is caused by the existence of two types of superexchange paths; one is ferromagnetic and the other is antiferromagnetic, canceling each other, and resultantly, $J_3$ becomes dominant \cite{CoplanarOct_Boldrin_prl_18,Sr-vesignieite_Boldrin_jmc_15, Ba-vesignieite_Boldrin_jmc_16}. Thus, in general, kagome antiferromagnets possessing several inequivalent superexchange paths for whatever reasons could be $J_3$-dominant candidates. The existence of an additional easy-axis anisotropy would be favorable for the occurrence of the collinear 12-sublattice state, as it can counteract the Dzyaloshinskii-Moriya interactions inherent to kagome systems that tend to cant the spins \cite{Sr-vesignieite_DM_Verrier_prb_20}. Once such a kagome antiferromagnet can be fabricated, it should serve as the emergent spin current generator.
 
\begin{acknowledgments}
The author (K. A.) thanks T. Aoyama for useful comments. We are thankful to ISSP, the University of Tokyo and YITP, Kyoto University for providing us with CPU time. This work is supported by JSPS KAKENHI Grant No. JP23H00257 and JP24K00572.
\end{acknowledgments}

\appendix
\section{Matrix elements of the Bgoliubov magnon Hamiltonian (\ref{eq:Hamiltonian_magnon})}
In the $J_1$-$J_2$-$J_2^\prime$ model on the square lattice [Eq. (\ref{eq:Hamiltonian_s_alter})], the ground state at $H=0$ is the 2-sublattice antiferromagnetic order where the sublattice moments can be expressed as ${\bf m}=(1,-1)$, namely, the sublattices 1 and 2 possess $\uparrow$ and $\downarrow$ spins, respectively. By using the Holstein-Primakoff transformation for the 2-sublattice order and expanding the Hamiltonian (\ref{eq:Hamiltonian_s_alter}) with respect to $1/S$, one obtains the Bogoliubov magnon Hamiltonian of the form ${\cal H}_{\rm alt} \simeq \frac{S}{2} \sum_{\bf q} \mbox{\boldmath $\Phi$}_{B{\bf q}}^\dagger  
 \left(\begin{array}{cc}
A_{\bf q} & B_{\bf q} \nonumber\\
B_{{\bf q}} & A_{{\bf q}} \nonumber
\end{array} \right) \mbox{\boldmath $\Phi$}_{B{\bf q}}$, where 2$\times$2 matrices $A_{\bf q}$ and $B_{\bf q}$ are given by
\begin{eqnarray}\label{eq:alt_coefficient}
&& A_{\bf q} = \big[ 4J_1 -2 J_2  - 2 J_2^\prime  +2D \big] I_{2\times 2} - (H/S) M_{2\times 2} \nonumber\\
&& \quad \, + \left( \begin{array}{cc}
J_2 & 0 \\
0 &  J_2^\prime 
\end{array}\right)  2 \cos({\bf q}\cdot {\bf d}_1 )  + \left( \begin{array}{cc}
J_2^\prime & 0 \\
0 &  J_2 
\end{array}\right) 2 \cos({\bf q}\cdot {\bf d}_3) , \nonumber\\
&& B_{\bf q} = 2J_1\big[ \cos({\bf q}\cdot {\bf a}_1) + \cos({\bf q}\cdot {\bf a}_3) \big] \left( \begin{array}{cc}
0 & 1 \\
1 &  0 
\end{array}\right),
\end{eqnarray}
where $I_{N\times N}$ is the $N \times N$ unit matrix and $(M_{N\times N})_{ij} =(I_{N\times N})_{ij} (-m_j)$. $A_{\bf q}$ and $B_{\bf q}$ in Eq. (\ref{eq:alt_coefficient}) are essentially the same as those given in Ref. \cite{alter_model_Brekke_prb_23}.

In the $J_1$-$J_3$ model on the kagome lattice [Eq. (\ref{eq:Hamiltonian_s_cfr})], on the other hand, the ground state at $H=0$ is the 12-sublattice collinear antiferromagnetic order where the sublattice moments are expressed as ${\bf m}=(-1,-1,-1,1,1,-1,1,-1,1,-1,1,1)$. In this case, $A_{\bf q}$ and $B_{\bf q}$ are 12$\times$12 matrices. In the same manner as that for Eq. (\ref{eq:alt_coefficient}), we obtain 
\begin{eqnarray}\label{eq:cfr_coefficient}
&& A_{\bf q} = (4J_3+2D) I_{12\times 12} - (H/S) M_{12 \times 12} \nonumber\\
&&\qquad +\left( \begin{array}{cccc}
D_0 & 0 & 0 & 0 \\
0 & D_1 & D_2^\ast & D_3^\ast \\
0 & D_2^\ast & D_2 & D_1^\ast \\
0 & D_3^\ast & D_1^\ast & D_3 
\end{array}\right), \nonumber\\
&& B_{\bf q} = \left( \begin{array}{cccc}
0 & C_1+D_1^\ast & C_3+D_3^\ast & C_2+D_2^\ast \\
C_1+D_1^\ast & D_2+D_3 & C_2 & C_3 \\
C_3+D_2^\ast & C_2 & D_1+D_2 & C_1 \\
C_2+D_2^\ast & C_3 & C_1 & D_1+D_3 
\end{array}\right),  \nonumber\\
&& D_l = J_1 \left( \begin{array}{ccc}
0& G_{l1,{\bf q}} &  G_{l3,{\bf q}}^\ast  \nonumber\\
G_{l1,{\bf q}}^\ast & 0 & G_{l2,{\bf q}} \nonumber\\
G_{l3,{\bf q}} & G_{l2,{\bf q}}^\ast & 0  \nonumber
\end{array}\right), \nonumber\\
&&C_l = J_3 \left( \begin{array}{ccc}
T_{l1,{\bf q}} + T_{l3,{\bf q}} &0 &  0 \nonumber\\
0 & T_{l1,{\bf q}} + T_{l2,{\bf q}} & 0 \nonumber\\
0 &0 & T_{l2,{\bf q}} + T_{l3,{\bf q}}  \nonumber
\end{array}\right), \nonumber\\
&& G_{lm, {\bf q}} =  s_{lm} \, e^{i{\bf q}\cdot {\bf e}_m}, \quad T_{lm,{\bf q}} = s_{lm} \, 2 \cos(2{\bf q}\cdot {\bf e}_m), \nonumber\\
&& s_{lm} = \delta_{l m}+\delta_{l 0} 
\end{eqnarray} 
with the real-space unit vectors connecting the NN pairs ${\bf e}_1=(1,0)$, ${\bf e}_2=(-\frac{1}{2}, \frac{\sqrt{3}}{2})$, and ${\bf e}_3=(-\frac{1}{2},-\frac{\sqrt{3}}{2})$.

\section{Details of the Monte Carlo simulation}
The temperature and magnetic-field phase diagrams for the Hamiltonians (\ref{eq:Hamiltonian_s_alter}) and (\ref{eq:Hamiltonian_s_cfr}) are obtained by performing MC simulations. In our MC simulations, $3\times 10^5$-$2\times 10^6$ MC sweeps are carried out under the periodic boundary condition, and the first half is discarded for thermalization, where a single spin-flip at each site consists of the conventional Metropolis update and a successive over-relaxation-like process in which we try to rotate a spin by the angle $\pi$ around the local mean field \cite{Loop_Shinaoka_14, SiteH_AK_21}. Observations are done at every MC sweep, and the statistical average is taken over 4 independent runs. The total number of spins $N$ is related with a linear system size $L$ via $N=L^2$ and $N=3L^2$ for the square-lattice [Eq. (\ref{eq:Hamiltonian_s_alter})] and kagome-lattice [Eq. (\ref{eq:Hamiltonian_s_cfr})] systems, respectively. 

By measuring various physical quantities such as the specific heat $C$, a Fourier magnetization as an order parameter $O_{\perp,\parallel}$, and the associated spin correlation length $\xi_{\perp,\parallel}$, we identify low-temperature phases. Here, the Fourier magnetizations are defined as the intensity of the spin structure factors 
\begin{eqnarray}
F_\perp({\bf q}) &=& \big\langle \big|\frac{1}{L^2} \sum_{\alpha = x,y }\sum_i S^\alpha_{{\bf r}_i} \, e^{i \, {\bf q}\cdot {\bf r}_i}\big|^2 \big\rangle,\nonumber\\
F_\parallel({\bf q}) &=& \big\langle \big|\frac{1}{L^2}\sum_i S^z_{{\bf r}_i} \, e^{i \, {\bf q}\cdot {\bf r}_i}\big|^2 \big\rangle,
\end{eqnarray}
at the ordering vector of ${\bf q}={\bf Q}$. Since due to the easy-axis anisotropy, ordering properties of the $S^z$ and $S^{x,y}$ components of the spin are different from each other, we treat these components separately. In the square-lattice case of Eq.(\ref{eq:Hamiltonian_s_alter}) where the conventional 2 sublattice antiferromagnetism with ${\bf Q} = (\pi, \pi)$ develops, the above physical quantities are given by \cite{trans-sq_AK_prb_19}
\begin{eqnarray}\label{eq:physq_alt}
C &=& \frac{1}{T^2 \, N}\Big( \langle {\cal H}_{\rm alt}^2 \rangle - \langle {\cal H}_{\rm alt} \rangle^2 \Big), \nonumber\\
O_{\perp,\parallel} &=& \sqrt{ F_{\perp,\parallel}({\bf Q}) }, \nonumber\\
\xi_{\perp,\parallel} &=& \frac{1}{|\mbox{\boldmath $\delta$}|} \sqrt{\frac{F_{\perp,\parallel} ({\bf Q})}{F_{\perp,\parallel}({\bf Q}+\mbox{\boldmath $\delta$})}-1 }
\end{eqnarray}
with $ \mbox{\boldmath $\delta$}= (\frac{2\pi}{L},0)$.
 
 In the kagome-lattice case of Eq. (\ref{eq:Hamiltonian_s_cfr}), the ordered phases are basically triple-${\bf Q}$ states, so that we define the order parameter $O_{\perp,\parallel}$ as the average of the spin structure factors over the three ${\bf Q}_\mu$'s, also defining the averaged spin correlation length $\xi_{\perp,\parallel}$ as follows \cite{KagomeSkX_AK_23}: 
\begin{eqnarray} \label{eq:physq_cfr}
O_{\perp,\parallel} &=& \sqrt{ \frac{1}{3}\sum_{\mu=1,2,3} F_{\perp,\parallel}({\bf Q}_\mu) }, \nonumber\\
\xi_{\perp, \parallel} &=& \frac{1}{3}\sum_{\mu=1,2,3}\frac{1}{|\mbox{\boldmath $\delta$}_\mu|} \sqrt{\frac{F_{\perp,\parallel}({\bf Q}_\mu)}{F_{\perp,\parallel}({\bf Q}_\mu+\mbox{\boldmath $\delta$}_\mu)}-1 }
\end{eqnarray}
with $\mbox{\boldmath $\delta$}_\mu=\frac{2\pi}{\sqrt{3} \, L}\hat{Q}_\mu$. 

\begin{figure}[t]
\begin{center}
\includegraphics[width=\columnwidth]{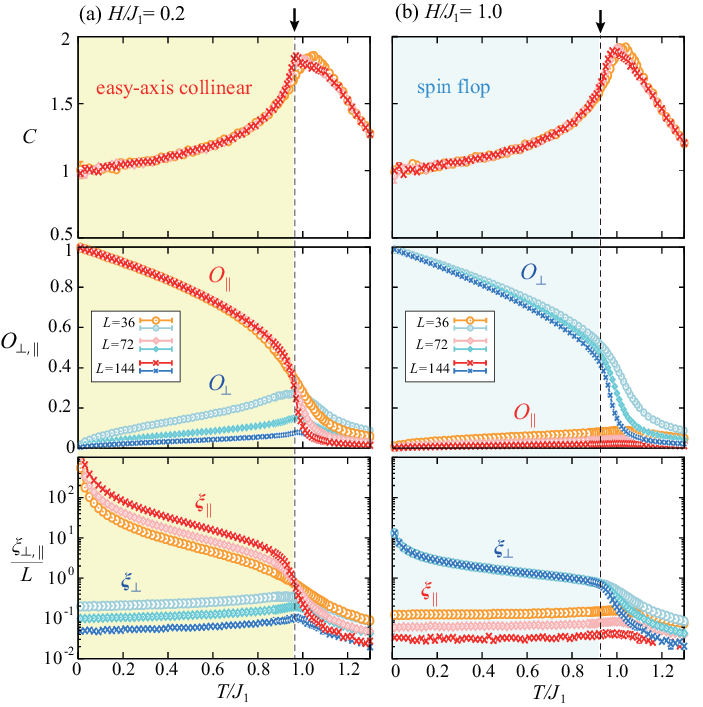}
\caption{MC results for the Hamiltonian (\ref{eq:Hamiltonian_s_alter}) on the square lattice with $J_2/J_1=-0.5$, $J_2^\prime/J_1=-0.25$, and $D/J_1=0.02$. (a) and (b) Temperature dependence of various physical quantities obtained in (a) the easy-axis collinear state at $H/J_1=0.2$ and (b) the spin flop state at $H/J_1=1.0$, where colored regions correspond to the phases in Fig. \ref{fig2} and black arrows indicate transition temperatures. Top panels: the specific heat $C$. Middle panels: the Fourier magnetizations for the $S^{x,y}$ and $S^z$ components $O_{\perp}$ and $O_{\parallel}$. Bottom panels: the associated spin correlation length ratios $\xi_\perp/L$ and $\xi_\parallel/L$. In the middle and bottom panels, bluish and reddish symbols correspond to the $S^{x,y}$ and $S^z$ components, respectively. \label{fig9} }
\end{center}
\end{figure}

In the present two dimensional systems with the easy-axis anisotropy, the easy-axis $z$ component of the spin $S^z$ can exhibit a long-range-order via a magnetic transition, whereas the in-plane $x$ and $y$ components $S^{x,y}$ can exhibit a quasi-long-range order via the Kosterlitz-Thouless (KT) transition. In the MC simulation, the former transition temperature can be identified as a point at which the correlation length ratios $\xi_\parallel/L$ obtained for various $L$'s cross, whereas the latter as a point at which the correlation length ratios $\xi_\perp/L$ for different sizes start merging  \cite{KT_Cuccoli_prb_95, trans-sq_AK_prb_19}. 

\begin{figure*}[t]
\begin{center}
\includegraphics[scale=0.8]{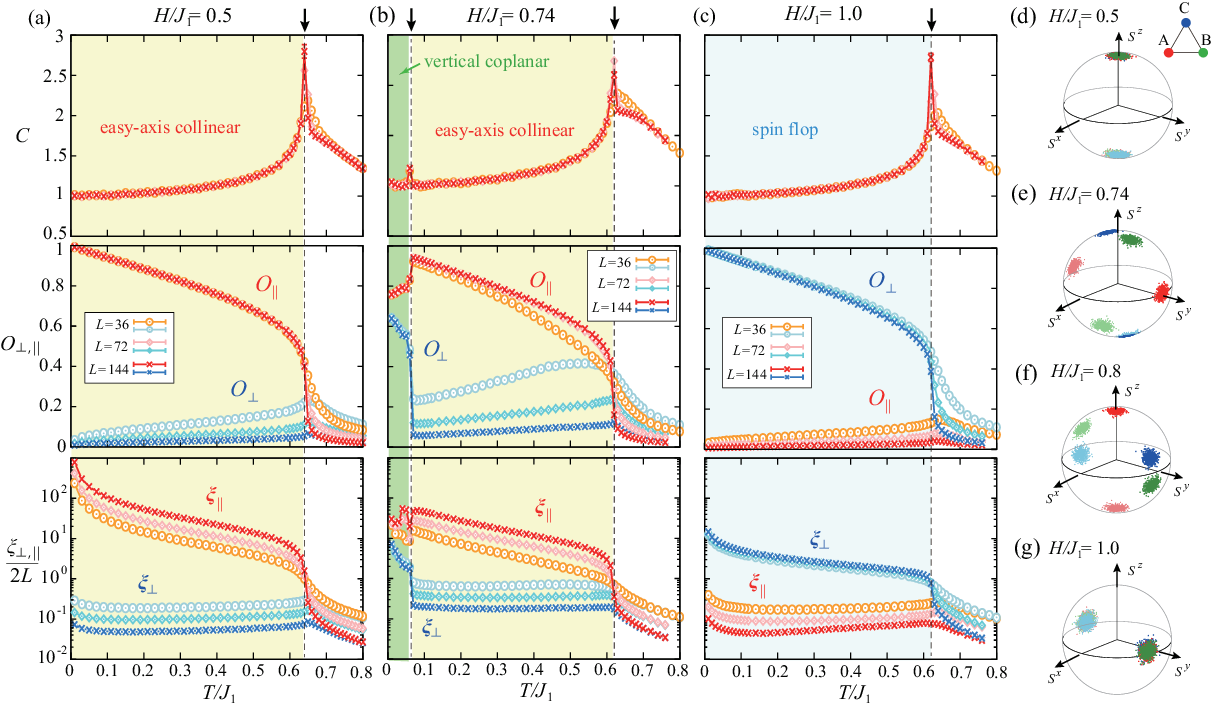}
\caption{MC results for the Hamiltonian (\ref{eq:Hamiltonian_s_cfr}) on the kagome lattice with $J_3/J_1=1.2$ and $D/J_1 = 0.02$. (a)-(c) Temperature dependence of various physical quantities obtained at (a) $H/J_1=0.5$, (b) 0.74, and (c) 1.0, where colored regions correspond to the phases in Fig. \ref{fig6} and black arrows indicate transition temperatures. Top panels: the specific heat $C$. Middle panels: the averaged Fourier magnetizations for the $S^{x,y}$ and $S^z$ components $O_\perp$ and $O_\parallel$. Bottom panels: the associated spin correlation length ratios $\xi_\perp/L$ and $\xi_\parallel/L$. In the middle and bottom panels, bluish and reddish symbols correspond to the $S^{x,y}$ and $S^z$ components, respectively. (d)-(g) Spin
snapshots mapped onto a unit sphere obtained in (d) the easy-axis collinear state at $H/J_1=0.5$, the vertical coplanar states at (e) $H_/J_1=0.74$ and (f) 0.8, and (g) the spin flop state at $H/J_1=1.0$, where the temperature is fixed to be $T/J_1=0.01$. Red, green, and blue points represent spins at the three corners of the upward triangle shown in the inset of (d). \label{fig10} }
\end{center}
\end{figure*}

Figure \ref{fig9} shows the temperature dependence of $C$ (top panels), $O_{\perp,\parallel}$ (middle panels), and $\xi_{\perp,\parallel}/L$ (bottom panels) at (a) $H/J_1=0.2$ and (b)$H/J_1=1.0$ in the square-lattice system of Eq. (\ref{eq:Hamiltonian_s_alter}). One can see from Fig. \ref{fig9} (a) that as indicated by a black arrow, the specific heat $C$ exhibits a peak at $T/J_1=0.97$ below which the $S^z$ component of the order parameter $O_\parallel$ takes nonzero values, suggestive of a phase transition into the 2 sublattice antiferromagnetic long-range-order in the $S^z$ sector. Actually, the associated correlation length ratios $\xi_\parallel/L$ for different $L$'s cross at this temperature [see reddish data in the bottom panel of Fig. \ref{fig9} (a)]. Note that the $S^{x,y}$ component remains disordered, as $O_\perp$ approaches zero in the thermodynamic limit of $L\rightarrow \infty$. At higher fields, on the other hand, as shown in Fig. \ref{fig9} (b), $C$ exhibits only a broad peak around $T/J_1=1.0$ and at a slightly lower temperature around $T/J_1 \sim 0.9$, $\xi_\perp/L$'s for different $L$'s start merging. Although $O_\perp$ might look nonzero, it gets gradually smaller with increasing $L$. These behaviors are typical of the KT transition whose temperature is indicated by the black arrow in Fig. \ref{fig9} (b). Since $O_\parallel$ tends to vanish in the $L\rightarrow 0$ limit, the antiferromagnetism develops only in the in-plane $S^{x,y}$ sector, and the $S^z$ component simply tends to align along the field. Thus, in this high-field phase, the spin-flop state is realized with the in-plane spin components exhibiting the quasi-long-range 2-sublattice order.

The above ordering features characteristic of the easy-axis antiferromagnets in two dimensions can also be seen in the kagome-lattice system of Eq. (\ref{eq:Hamiltonian_s_cfr}). Figures \ref{fig10} (a) and (c) show low-field and high-field data, respectively. One can see that the $S^z$ and $S^{x,y}$ components exhibit a long-range order at low fields and a quasi long-range order at high fields, respectively, with their counter components remaining disordered. 
Compared with the square-lattice case, $C$ in the kagome-lattice case shows a relatively sharp peak (compare the top panels in Figs. \ref{fig9} and \ref{fig10}), which could be attributed to an additional bond or Potts ordering originating from the triple-${\bf Q}$ nature (see \cite{RMO-collinear_Grison_prb_20} and supplemental material in \cite{KagomeSkX_AK_22}). 
MC spin snapshots in the low-field and high-field phases are shown in Figs. \ref{fig10} (d) and (g), respectively, where three corner spins on each upward triangle, i.e., spins belonging to the corners A, B, and C [see the inset of Fig. \ref{fig10} (d)], are represented by red, green, and blue points on a sphere, respectively. It can clearly be seen that the easy-axis collinear (spin-flop) state is realized in the low-field (high-field) phase. Note that in both the low-field and high-field phases, the corner spins, A, B, and C, exhibit the same ordering behavior, as their data points overlap on the sphere [see Figs. \ref{fig10} (d) and (g)].

In a narrow field range between the low-field easy-axis collinear and high-field spin-flop phases, there appear a phase in which the A-, B-, and C-sublattice spins behave in a different manner. As shown in Fig. \ref{fig10} (e), at $H/J_1=0.74$ just above the low-field easy-axis collinear phase, the B- and C-sublattice spins are slightly tilted from the easy axis, whereas the A-sublattice spins exhibit a spin-flop structure, constituting a vertical coplanar structure as a whole. This vertical coplanar phase corresponds to the lowest temperature phase in Fig. \ref{fig10} (b) where both the long-range-order of the $S^z$ component and the quasi-long-range order of the $S^{x,y}$ component develop (see the order parameter $O_\parallel$ in the middle panel and the correlation length ratio $\xi_\perp/(2L)$ in the bottom panel). At the slightly higher field of $H/J_1=0.8$ [see Fig. \ref{fig10} (f)], the A, B, and C spins are rotated roughly by 90$^\circ$ in the coplanar plane with the whole spin structure remaining the vertical coplanar. Although these two types of coplanar states might be distinguishable from a symmetry consideration, we simply dub these spin structures the vertical coplanar.

\end{document}